\documentstyle[12pt]{article}
\def\singlespace {\smallskipamount=3.75pt plus1pt minus1pt
                  \medskipamount=7.5pt plus2pt minus2pt
                  \bigskipamount=15pt plus4pt minus4pt
                  \normalbaselineskip=15pt plus0pt minus0pt
                  \normallineskip=1pt
                  \normallineskiplimit=0pt
                  \jot=3.75pt
                  {\def\smallskip {\vskip\smallskipamount}}
                  {\def\medskip   {\vskip\medskipamount}}
                  {\def\bigskip   {\vskip\bigskipamount}}
                  {\setbox\strutbox=\hbox{\vrule
                    height10.5pt depth4.5pt width 0pt}}
                  \parskip 7.5pt
                  \normalbaselines}
\def\middlespace {\smallskipamount=5.825pt plus1.5pt minus1.5pt
                  \medskipamount=11.25pt plus3pt minus3pt
                  \bigskipamount=22.5pt plus6pt minus6pt
                  \normalbaselineskip=22.5pt plus0pt minus0pt
                  \normallineskip=1pt
                  \normallineskiplimit=0pt
                  \jot=5.825pt
                  {\def\smallskip {\vskip\smallskipamount}}
                  {\def\medskip   {\vskip\medskipamount}}
                  {\def\bigskip   {\vskip\bigskipamount}}
                  {\setbox\strutbox=\hbox{\vrule
                    height15.75pt depth6.75pt width 0pt}}
                  \parskip 7.25pt
                  \normalbaselines}
\def\dblspc {\smallskipamount=7.5pt plus2pt minus2pt
                  \medskipamount=15pt plus4pt minus4pt
                  \bigskipamount=30pt plus8pt minus8pt
                  \normalbaselineskip=30pt plus0pt minus0pt
                  \normallineskip=2pt
                  \normallineskiplimit=0pt
                  \jot=7.5pt
                  {\def\smallskip {\vskip\smallskipamount}}
                  {\def\medskip   {\vskip\medskipamount}}
                  {\def\bigskip   {\vskip\bigskipamount}}
                  {\setbox\strutbox=\hbox{\vrule
                    height21.0pt depth9.0pt width 0pt}}
                  \parskip 15.0pt
                  \normalbaselines}
\def\nb{\nabla }

\def\la {\lambda }

\def\be{\begin{equation}}

\def\j-{\J_-}

\def\eps{\epsilon}
\def\gm{\gamma}

\def\ee{\end{equation}}

\def\bea{\begin{eqnarray}}
\def\bearr{\begin{eqnarray}}
\def\bearrs{\begin{eqnarray*}}
\def\eearr{\end{eqnarray}}
\def\eea{\end{eqnarray}}
\def\eearrs{\end{eqnarray*}}
\def\barr{\begin{array}}
\def\earr{\end{array}}
\def\p{\partial}

\def\th{\theta}

\def\o{\omega}

\def\non\non{\nonumber}
\def\nn8{\nonumber\\[15pt]}
\def\l{\left}
\def\r{\right}
\def\un{\underline}

\def\f{\frac}

\begin{document}
\middlespace
\begin{center}
{\Large{\bf General Relativistic Contribution to
Polarization\\[6pt] of the Cosmic Microwave Background
Anisotropy}}\\[12pt]
{\bf S. Mohanty and A.R. Prasanna\\ Physical Research Laboratory\\
Ahmedabad 380 009, India}\\[12pt]

\un{Abstract}\\
\end{center}
We solve the wave equation for the electromagnetic field tensors
associated with the CMBR photons in a universe with scalar metric
perturbations. We show that the coupling of the electromagnetic
fields with the curvature associated with the scalar perturbations
gives rise to an optical rotation of the microwave background
photons. The magnitude of the gravitationally generated V-Stokes
parameter anisotropy $\Delta_V$, is however very small compared to
the linear polarisation caused by Thomson scattering.\\ \newpage

In the standard treatment of the microwave background anisotropies
[1] it is assumed that the gravitational potential at the last
scattering surface gives rise to temperature anisotropies at large
angular scales through the Sachs-Wolfe effect [2].
Thomson scatterings near the surface of last scattering
cause a linear polarisation
of the CMB photons. 

 The anisotropies of CMBR are governed
by the
the following
set of equations:
 The Boltzmann equation for the photon ditribution function $f(p_i,
p_0,z)$
\be
{Df \over D \lambda}=p^0 {\partial f\over  \partial x^0} +
p^i {\partial f\over  \partial x^i}+ {d p^i \over d \lambda}
{\partial f\over  \partial p^i} = 0
\ee
the geodesic equation of photons in the perturbed metric
\be
{dp^\alpha\over d \lambda} + {\Gamma^\alpha}_{\beta \gamma}
p^{\beta}p^{\gamma}=0
\ee

and  the photon disperison relation
\be
 g_{\mu \nu} p^{\mu} p^{\nu}=0.
\ee

The dispersion relation (3) is derived in the zeroeth order of eikonal
approximation from the electromagnetic field wave equation in a
curved
spacetime.
In the standard
eikonal approximation of photon trajectories in gravitational
fields, the red shift of the photon is independent of the
polarization of its spin and gravity
therefore makes no contribution to the polarization of the CMB
radiation fields.\\

It has been shown in [3]that the gauge invariant wave equation of
electromagnetic fields in curved spacetimes is given by 
\be
\nb^\mu \nb_\mu F_{\nu\la} + R_{\rho \mu \nu \la} \;F^{\rho \mu} +
R^\rho_{\;\;\la} F_{\nu \rho} - R^\rho_{\;\; \la} F_{\la \rho} =0
\label{wave}
\ee
 Electromagnetic field tensors couple to the Riemann and Ricci
curvature tensors of the background spacetime and these couplings
can give rise to polarisation of electromagnetic waves in
gravitational fields. In this we show that scalar perturbations to
the Robertson-Walker metric near the surface of last scattering
can give rise to circular polarization to the CMB radiation. We
calculate the anisotropy in this gravitationally generated
circular polarisation as a function of angular separation of the
beams. We find however that the magnitude of this circular polarization
anisotropy is unlikely to be observed in the planned CMB
polarisation experiments [4].\\

We consider the scalar perturbations in the Friedmann- Robertson-Walker
 Universe
described by the metric [5]
\be ds^2 = a^2(\eta) \l[ (1 + 2 \psi)
d\eta^2 - (1 - 2\phi ) \gm_{ij} dx^i dx^j \r]
\label{metric}
\ee
In the synchronous
gauge the two scalar perturbation are the Newtonian potential
$\psi$ and the spatial curvature $\phi$. The eikonal equation
obeyed by the photon wave vector $q_\mu = \l( q_o,
\vec{q} \r)$ which follows from the wave equation (1) is given by
[3]
\be \l( q^2 \cdot \delta^i_{\;\;j} + \eps^i_{\;\; j}\r) f_{oi} =0
\label{disp}
\ee
where $f_{oi}$ is the amplitude of the electric field
associated with the radiation and the anisotropic gravitational
permeability tensor $\eps^i_{\;\;j}$ is given by [3]
\be
\barr{lll} \eps^j_{\;\;i} = \l( R^o_{\;\;o} + R^\ell_{\;\; o}
~~\f{q_\ell}{q_o} \r) \delta^j_{\;\;i}+ \l( - 2
R^{oj}_{\;\;\; oi} + 4 R^{\ell j}_{\;\;\; oi}~~ \f{q_\ell}{q_o} +
R^j_{\;\;i} - R^j_{\;\;o}~~ \f{q_i}{q_o}\r) \earr
\label{eij}
 \ee
 The components of
the permeability tensor $\eps^j_{\;\;i}$ which follows from the
metric (2) are given in Appendix A.

Consider a photon wavenumber $q$ and choose the z-axis along
$\vec{q}$. The wave equations for the transverse $\vec{E}$ field
components is
\be
\l( \barr{cc} q^2 + \eps^1_{\;\;1}& \eps^2_{\;\;1}\\
\eps^1_{\;\;2}& q^2 + \eps^2_{\;\;2} \earr \r) \l( \barr{c} E_1
\\E_2 \earr \r) = 0
\ee
where
\bea
 \eps^1_{\;\;1} &=& - \f{8\pi G}{3} \bar{\rho} ( 1 - 2
\psi ) + \f{2}{a^2} \l( k^1 k_1 \r) \psi \l( \vec{k}, \eta \r)
  \\
\eps^2_1 &=& \f{2}{a^2} k^2 k_1 \psi \l( \vec{k},
\eta \r)
 \eea
 and similarly for the
others, where $\vec{k} = ( k_1, k_2, k_3 )$ is the wavenumber
of the gradients of the scalar perturbation $\psi ( \vec{k},
\eta )$. The dispersion relations of the two
propagating modes $E_\pm$ are obtained by setting the determinant
of the matrix operator (5) to zero to give
\be
\barr{lll} q^2_{0\; \pm} = q^2 \l[ 1 - \f{\eps^1_{\;\;1} +
\eps^2_{\;\;2}}{2q^2} \pm \f{1}{2q^2} \l(
\l( \eps^1_{\;\;1} - \eps^2_{\;\;2} \r)^2 + 4 \eps^1_{\;\;2}
\eps^2_{\;\;1} \r)^{1/2} \r] \earr
\ee
 The eigenvectors $E_\pm$
are given in terms of the mixing angle $\chi$ as
\be
\l( \barr{c} E_+\\ E_- \earr \r) = \l( \barr{cc} \cos
\chi&\sin\chi\\ - \sin\chi&\cos\chi \earr \r) \l( \barr{c}
E_1\\E_2\earr \r)
\label{eigen}
\ee where the mixing angle $\chi$ is given by
\be
\tan 2\chi = \f{2 \eps^1_{\;\;2} \eps^2_{\;\;1}}{\l(
\eps^1_{\;\;1} - \eps^2_{\;\;2} \r)}
\label{chi}
\ee
 Expressing
$\eps^i_{\;\;j}$ in terms of polar coordinates of $\vec{k}$ , we have
\be
q_{0\pm} = q [ 1+ {8\pi G \over 3 q^2} \bar\rho (1-2\psi)
\mp {k^2 sin^2 \theta~ \psi \over q^2 a^2}  ]^{1/2}
\ee
 and the mixing angle as defined in (\ref{chi})  is
 \be \tan 2\chi =
\tan 2\phi \ee
 The normal mode solutions $E_\pm$ are therefore
given by
\be
\barr{lll} E_+&=& \l[ E_1 \cos \phi + E_2 \sin \phi \r]e^{i\l(
(q_{0+}  \eta) - qz \r)}\\[8pt]
 E_-&=& \l[ - E_1 \sin\phi + E_2
\cos\phi \r] e^{i\l( (q_{0 -} \eta) - qz \r)}
\earr
\label{epm}
\ee
The Stokes
parameters of a radiation field are described by the $2\times 2$
complex matrix [5],
\be
\rho_{ij} = \f{\left <E_i E_j^{*} \right>}{I} = \f{1}{2} \l(
\barr{cc} I+Q&U-iV\\U+iV&I-Q\earr \r) \ee
 where
\bea
\barr{lll}
 I \equiv \left <
E_+ E_+^{\;\;*} + E_-E^{\;\;*} \right > \\
Q \equiv
\left < E_+ E_+^{\;\;*} - E_-E^{\;\;*} \right > \\
U
\equiv \left < E_+E_-^{\;\;*} + E_- E_+^{\;\;*} \right >;\\
 V = i \left < E_+ E_-^{\;\;*} - E_+^{\;\;*} E_-
\right > \\
\earr
\label{stokes}
\eea
 where the time average is taken over duration
larger than the inverse frequency.  In unpolarised radiation $Q = U
= V = 0$. The degree of polarisation is defined by a vector
$\vec{P}$ with magnitude $P = \sqrt{Q^2 + U^2 + V^2}$. The
quantity $\ell = \sqrt{Q^2 + U^2}$ denotes the degree of linear
polarization and $V$ denotes the circular polarisation of the
radiation. Using the eigenmodes in the gravitational field given
by (\ref{epm}) and using the definitions (\ref{stokes}) we find the 
Stokes
parameters of the propagating beam are
\bea
\barr{lll}
 Q(\eta) = Q\l( \eta_i \r) \cos 2\phi + U(\eta_i) sin 2 \phi \\[8pt]
U(\eta ) =  U\l( \eta_i \r) \cos
2\phi ~cos \delta (\eta-\eta_i)- Q \l( \eta_i \r) \sin 2\phi
\cos ( \Delta \omega(\eta - \eta_i )) \\[8pt]

V(\eta ) = \l(- U \l( \eta_i
\r) \cos 2 \phi + Q \l( \eta_i \r) \sin 2\phi \r) \sin ( \Delta \omega
( \eta - \eta_i ) ) \\[8pt]
  I\l( \eta \r) = I \l(
\eta_i \r)
\earr
\eea
 where
\be
\Delta \omega \equiv \l( q_{0+} -
q_{0-} \r) = \f{ k^2}{a^2 q} \sin^2\th  \psi ( k, \eta)
\ee
is the rate at which the plane of polarisation is rotated.

We see that if at some initial time $\eta_i$ there is non-zero $Q
\l( \eta_i \r)$ or $U \l( \eta_i \r)$ then at a later time a
circular polarization $V(\eta )$ is generated.  The intensity of
the beam $I(\eta )$ remains unchanged. One can express (16)
as differential equations for evolution of $Q,U,V$ as

\bea
\barr{lll}
 \f{\p U}{\p\eta} =  (\Delta \o ) \sin \l( \Delta \o \eta \r) \l[ U 
\cos
2\phi - Q \sin 2\phi \r] \\

\f{\p Q}{\p \eta} = 0\\

 \f{\p V}{\p \eta} = \l( \Delta \o \r) \cos \l(
\Delta \o \eta \r) \l[- U \cos 2 \phi + Q \sin 2 \phi \r]
\earr
\label{stokes3}
\eea
 To the leading order in
$(\Delta \o )$ we see that $\dot{U} = \dot{Q} = 0$ and
\be
\f{\p V}{\p \eta} = (\Delta \o ) \l[ U \cos 2 \phi - Q \sin 2 \phi \r]
\ee
Assuming that at the time of decoupling $\eta_i$ there is a
non-zero $U \l( \eta_i \r) \; Q \l( \eta_i \r)$ due to Thomson
scattering, one can estimate the degree of circular polarisiation
$V$ in CMB radiation.

From the form of equation (\ref{stokes3}) it is clear that the circular
 polarisation
$V$ averaged over all angles $\phi$ of the gravitational perturbation 
vector
$\vec k$ will be zero. The rms value of $V$ defined as $\Delta_V
(q,k,cos \theta) =  <(V - <V>)^2>^{1/2} $. Similarly defining
$\Delta_Q$
and,
$\Delta_U$) , we see that
 the polarization anisotropies obey the coupled set of
Boltzmann equations [6],
\be
\dot{\Delta}_U + ik\mu \Delta_U = - \dot{\kappa} \Delta U
\label{du}
 \ee
\be
\dot{\Delta}_Q + ik\mu \Delta_Q = - \dot{\kappa} \l[ \Delta_Q -
\f{1}{2} \l( 1 - P_2 (\mu ) \r) S_P \r]
\label{dq}
 \ee
\be
\dot{\Delta}_V + ik\mu \Delta_V = \l(
\Delta_Q \sin 2\phi \r) (\Delta \o )
\label{dv}
 \ee
where $\mu = \cos \th$
and $S_P = - \l( \f{5}{2} \r) \Delta T_2$ the quadrupole
temperature anisotropy.  The gravitational contribution arises as
a source term for the $V$-polarisation mode while the $Q$ and $U$
modes are generated by Thomson scattering parametrised by $\dot{\kappa}
\equiv \l( \chi_e n_e \sigma_T a(\eta ) / a\l( \eta_o \r) \r)$
with $\chi_e$ the ionised fraction, $n_e$ the electron number
density and $\sigma_T$ the Thomson scattering cross section. In
the tight coupling regime (keeping only leading order terms
in $\dot{\kappa}^{-1}$ ) the solutions
of (\ref{du}) and (\ref{dq}) are given by
\bea
\Delta_U = 0 \\
 \Delta_Q = - \f{15}{8} \sin^2\th \Delta_{T_2}
\label{tc}
\eea where we see that the $Q$ polarization is generated by a
quadrupole temperature anisotropy $\Delta_{T_2}$.

Using (\ref{dv}) and (\ref{tc})
the solution of $\Delta_V \l( k,\eta , \mu \r)$  is
\be
\Delta_V \l( k, \eta , \mu \r) = e^{-i\mu \l( \eta - \eta_i \r)}
\l( \f{-15}{8} \r) \sin^2\th \Delta_{ T_2} \Delta \o \l( \eta -
\eta_i \r) \ee
One can estimate that the two point correlations
\be
C^{VT} ( \th ) \equiv \left < \Delta_T \l( \hat{n}_1 \r) \Delta_V
\l( \hat{n}_2 \r) \right >
\ee
and
\be
C^{TT} (\th ) \equiv \left < \Delta_T \l( \hat{n}_1 \r) \Delta_T
\l( \hat{n}_2 \r) \right >
\ee
 have relative magnitudes
\be
\f{C^{VT}(\th )}{C^{TT} (\th )} \simeq \Delta \o \cdot \l( \eta -
\eta_i \r) \simeq \f{k^2_{max}}{q} H^{-1}_o
 \ee
Taking the photons wavelength $\sim 1 m$ and
the smallest measurable metric perturbation to be the size of galaxies 
$\sim 100 kpc$, we find that the GTR contribution to polarisation
anisotrpy is smaller than the temperature anisotropy by a factor of
$10^{-41}$. This may be compared with the corresponding factor for the    
ratio of the polarization anisotropy due to Thomson scattering to
temperature anisotropy  which is $10^{-7}$.

This ratio is very
small unless the anisotropies are observed at very small angles
(large $K_{max}$) but here the observations are difficult because
of contamination from point sources. In conclusion, we note that
gravitational couplings of the scalar perturbations on photon
polarisations can generate circular polarisation anisotropies in
principle. However, in practice, it would be difficult to
observe.

\newpage
\begin{center}
{\bf Appendix}\\
\end{center}

For the metric (2) with $\phi = \psi$ Riemann and Ricci
components of curvature are
\[
\barr{lll} R^{ij}_{\;\;\;k\ell}&=& \f{1}{a^2} \l[ \psi^i_{\; \ell}
\delta^j_{\; k} - \psi^i_{\; k} \delta^j_{\; \ell} + \psi^j_{\; k}
\delta^i_{\; \ell} - \psi^j_{\; \ell} \delta^i_{\; k}\r] \\[8pt]
&=& - \f{\dot{a}}{a^3} \l[ \f{\dot{a}}{a} \l( 1 - 2\psi \r) - 2
\dot{\psi} \r] \l( \delta^i_{\; k} \delta^j_{\; \ell} -
\delta^i_{\; \ell} \delta^j_{\; k} \r)\\[8pt]
 R^{oi}_{\;\;\;oj}
&=&
\f{1}{a^2} \l\{ \psi^i_{\;j} - \delta^i_{\; j} \l[ \l(
\f{\ddot{a}}{a} - \f{\dot{a}^2}{a^2} \r) \l( 1 - 2 \psi \r) -
\f{2\dot{a}\dot{\psi}}{a} - \ddot{\psi} \r] \right.\\[8pt]
R^{ij}_{\;\;\;ok}&=& \f{1}{a^2} \l\{ \l( \dot{\psi}^j +
\f{\dot{a}}{a} \psi^j \r) \delta^i_{\;k} - \l( \dot{\psi}^i +
\f{\dot{a}}{a} \psi^i \r) \delta^j_{\;k} \r\}\\[8pt] R^i_{\;j}&=&
- \f{1}{a^2} \delta^i_{\;j} \l\{ \l( \f{\ddot{a}}{a} +
\f{\dot{a}^2}{a^2} \r) \l( 1 -2 \psi \r) - \ddot{\psi} -
\f{6\dot{a}\dot{\psi}}{a} + \psi^k_{\;k} \r\}\\[8pt] R^o_{\;o}&=&
\f{1}{a^2} \l\{ \psi^k_{\;k} - 3 \l[ \l( \f{\ddot{a}}{a} -
\f{\dot{a}^2}{a^2} \r) \l( 1 - 2\psi \r) -
\f{2\dot{a}\dot{\psi}}{a} - \ddot{\psi} \r] \right.\\[8pt]
R^i_{\;o}&=& - \f{2}{a^2} \l( \dot{\psi}^i + \f{\dot{a}}{a} \psi^i
\r) \earr
\]
and the field equations $R^i_{\; j} = 8\pi G \l( T^i_{\;j} -
\f{1}{2} \delta^i_{\; j} T \r)$ give
\[
\barr{rll} \psi^k_{\;k} - 3 \l[ \l( \f{\ddot{a}}{a} -
\f{\dot{a}^2}{a^2} \r) \l( 1 - 2\psi \r) -
\f{2\dot{a}\dot{\psi}}{a} - \ddot{\psi} \r] &=& 4\pi G \rho_b a^2
\l( 1 + 3 \nu \r) \l( 1 + \delta \rho \r)\\[8pt] \dot{\psi}^i +
\f{\dot{a}}{a} \psi^i &=& 4\pi G \l( \rho_b a^2 \r) \l( 1 + \nu
\r) V^i\\[8pt] \l( \f{\ddot{a}}{a} + \f{\dot{a}^2}{a^2} \r) \l( 1
- 2 \psi \r) - \ddot{\psi} - \f{6\dot{a}\dot{\psi}}{a} +
\psi^k_{\;k}&=& 4 \pi G \rho_b a^2 \l( 1 - \nu \r) \l( 1 + \delta
\rho \r) \earr
\]
wherein the equation of state $\rho = \f{p}{\nu}$ is used.\\

The $\eps$ matrix is then given by
\[
\barr{lll}
\eps^k_{\;k} &=& -\f{2}{a^2} \l\{ \psi^k_{\;k} + \f{\ddot{a}}{a}
\l( 1 - 2 \psi \r) - \ddot{\psi} - \f{4\dot{a}\dot{\psi}}{a}
\r\} \; \; \barr{c} \l( k = 1,2,3 \r)\\ \mbox{ no summation }
\earr \\[8pt]
\eps^1_{\;2}&=& - \f{2}{a^2} \psi^1_{\; 2}\\[8pt]
 \eps^2_{\; 1}&=&
- \f{2}{a^2} \psi^2_{\; 1}\\[8pt]
 \eps^3_{\; 1}&=& -\f{2}{a^2}
\psi^3_{\; 1}\\[8pt] \eps^3_{\; 2}&=& - \f{2}{a^2} \psi^3_{\;
2}\\[8pt] \eps^1_{\; 3}&=& - \f{2}{a^2} \l[ \psi^1_{\; 3} -
\f{K_3}{K_o} \l( \dot{\psi}^1 + \f{\dot{a}}{a} \psi^1 \r)
\r]\\[8pt] \eps^2_{\;3}&=& - \f{2}{a^2} \l[ \psi^2_{\; 3} -
\f{K_3}{K_o} \l( \dot{\psi}^2 + \f{\dot{a}}{a} \psi^2 \r) \r]
\earr
\]
\newpage
\begin{center}
{\bf References}\\ \end{center}
 \begin{enumerate}
\item W. Hu and N. Sugiyama, {\it Phys. Rev.} {\bf D51}, 2599 (1995).\\
{\it Astrophy. J.} {\bf 445} 521 (1995).
\item R.S. Sachs and A.M.Wolfe, {\it Ap. J.} {\bf 147} 73 (1967)
\item S. Mohanty and A.R. Prasanna, {\it Nucl. Phy.} {\bf B526}
501 (1998)
\item M. Zaldarriaga, D.N. Spergel and U. Seljak, {\it Ap. J.}
{\bf 488} 1 (1997)\\ MAP and Planck details in
http://map.gsfc.nasa.gov \\
http://map.gsfc.nasa.gov \\
http://astro-estec.esa.nl/SA-general/Projects/Planck\\

\item  J.M. Bardeen , Phys.Rev.{\bf D22} :1882-1905 (1980)\\ 
 L. Landau and E.M. Lifshitz, {\it The Classical Theory of
Fields}, Pergamon Press, London.
\item J.R. Bond and G. Efstathion, {\it Ap. J.} {\bf 285}, L45
(1984)\\ M. Zaldarriaga and D. Harari, {\it Phys. Rev.} {\bf D51}
2599 (1995)\\ D. Harari, J.D. Hayward and M. Zaldarriaga, {\it Ap.
J.} (1996)
\end{enumerate}
\end{document}